# Design and Early Development of a UAV Terminal and a Ground Station for Laser Communications

Alberto Carrasco-Casado*, Ricardo Vergaz, José M. Sánchez-Pena

Dpto. Tecnología Electrónica, Universidad Carlos III de Madrid
Avda. Universidad, 30, 28911, Madrid, Spain;

## ABSTRACT

A free-space laser communication system has been designed and partially developed as an alternative to standard RF links from UAV to ground stations. This project belongs to the SINTONIA program (acronym in Spanish for low environmental-impact unmanned systems), led by BR&TE (Boeing Research and Technology Europe) with the purpose of boosting Spanish UAV technology.

A MEMS-based modulating retroreflector has been proposed as a communication terminal onboard the UAV, allowing both the laser transmitter and the acquisition, tracking and pointing subsystems to be eliminated. This results in an important reduction of power, size and weight, moving the burden to the ground station. In the ground station, the ATP subsystem is based on a GPS-aided two-axis gimbal for tracking and coarse pointing, and a fast steering mirror for fine pointing. A beacon-based system has been designed, taking advantage of the retroreflector optical principle, in order to determine the position of the UAV in real-time. The system manages the laser power in an optimal way, based on a distance-dependent beam-divergence control and by creating two different optical paths within the same physical path using different states of polarization.

## 1. INTRODUCTION

A free-space optical communication system has been designed and partially developed as an alternative to standard radio frequency (RF) links from Unmanned Aerial Vehicles (UAV) to ground stations. This project belongs to the SINTONIA [1] program (acronym in Spanish for low-environmental-impact unmanned systems), led by BR&TE (Boeing Research and Technology Europe) with the purpose of boosting Spanish UAV technology. The work of GDAF-UC3M is under the coordination of INDRA, S.A. and INSA, S.A. companies.

Small UAVs are becoming a key part of national security and it is foreseen that in the future this tendency will continue with a stronger and stronger impact. A great flexibility has been demonstrated by these kinds of aircraft, which have been used in a big number of both civil and military objectives and scenarios, from agriculture to meteorology and from research to military warfare. The US Army alone holds over 4,000 UAVs with many more programmed [2].

Communications play a more important role in the operation of unmanned aircraft than they do in manned ones because all the decision-making occurs on the ground, either before or during the flight itself. Currently, telecommunications links between UAVs and ground stations are based on RF systems and low-earth-orbiting satellites [3]. Both are long range communications but also have low bit rates, usually in the order of hundreds of kbps or less, and in the case of satellite communications the payload involves an important burden regarding to mass and weight onboard.

The move to optical carrier frequencies involves a qualitative leap because it provides a shift of several orders of magnitude, from MHz to hundreds of thousands of GHz. Since the minimum divergence, given by the diffraction limit of an aperture, is dependent on the wavelength of the electromagnetic wave [4], this shift implies also a dramatic decrease of about five orders of magnitude in the divergence of the communication signal. Lower divergence allows higher reception power and signal-to-noise ratio, enabling faster communications with lower bit-error-rates [5].

*aacarras@ing.uc3m.es

Free-space lasercom is also a more secure technique than RF transmission, which has led to important vulnerabilities that became evident in the interception of the downlink of a US Predator by Iraqi fighters in 2009 [6]. The extremely high directivity of the optical communication links makes the laser beams extremely difficult to detect by other than the legitimate receivers and virtually impossible to intercept without being noticed since an interception leads to a signal fading.

Optical wavelengths also offer a solution for the bandwidth shortage, which limits the use of the number of UAVs in operation over a given geographical area because the electromagnetic spectrum is scarce. UAVs must operate in very crowded frequency and bandwidth spectrums and this problem will be more and more evident with the increasing addition of new terminals. A free-space lasercom system does not have to compete for electromagnetic broadcast bands in the spectrum at all. Another big advantage of this technology is that it enables lower power consumption and smaller and lighter terminals [7]. Since a first-class goal in the SINTONIA program is developing low-impact technology onboard UAVs, optical communications fit ideally these requirements.

Additionally, the use of optical communications with small UAVs will allow a natural integration with an upcoming important technology, i.e., laser power transmission to UAVs. This technique has the capability to provide energy to electric aircraft for an unlimited amount of time to dramatically extend mission run times. The ability to fly continuously without landing for weeks or months would be a breakthrough in aviation, with UAVs serving as "satellites" on surveillance or telecommunications relay missions [8].

## 2. REMOTE TERMINAL DESIGN

Given the strong effort needed to optimize the telecommunications payload, a modulating retroreflector (MRR) has been proposed as the communications remote terminal (Fig. 1). A retroreflector sends the incoming beam back to the ground station through the same path of the interrogator laser.

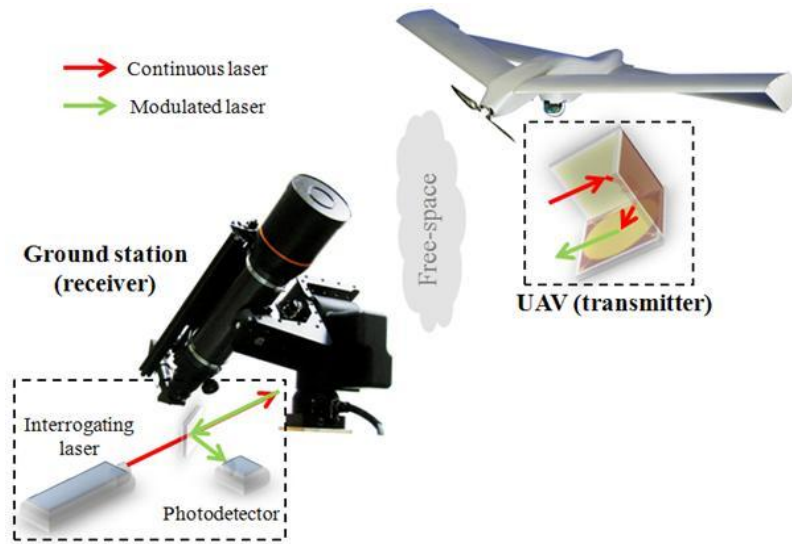

Figure 1. Modulating retroreflector principle.

This device combines a retro-reflector and an optical modulator and is capable of returning light from a distant interrogating laser source without any additional pointing requirement onboard, while simultaneously modulating its intensity on the way back [9]. Such a device allows both the laser transmitter subsystem and the acquisition, tracking and pointing (ATP) subsystem to be fully eliminated at one end of the link, which results in a considerable reduction of power, size and weight onboard the UAV. The burden moves to the ground station but the ATP subsystem is eased since the MRR acts as a pointing reference by reflecting the incoming laser beam back to its source. This technique is further explained in section 3.1.

Both the viability of lasercom with minimum communication payload and high-rate communications with UAVs are going to be demonstrated. The former will use a MEMS-based (MicroElectroMechanical Systems) MRR switched in the order of hundreds of kbps, a system that is expected to weigh less than 1 kilogram and take up a few centimeters, thus reducing payload. The latter will take advantage of an OOK-modulation of the ground laser transmitter in the order of hundreds of Mbps. The proposed system can allow optical communications on terminals that would otherwise not be able to support lasercom due to weight/mass/power constraints.

Sections 2.1 and 2.2 will show two alternatives to develop the modulating retroreflector.

## 2.1 Liquid crystal-based modulating retroreflector

In the past, experiments were made [11] by the authors of this paper testing liquid crystal technology as a transmissive modulator in an MRR scheme. Several configurations were tested, being the PolSK (Polarization Shift Keying) the chosen one. This technique offers the capability to modulate information as changes in the State Of Polarization (SOP) of light, allowing to create a big constellation of symbols that could be demodulated by a receiver consisting on parallel SOP measurements. V-shaped smectic liquid crystal was chosen as the active element in the MRR and 32 polarization levels were achieved increasing the data rate in a 5-fold factor.

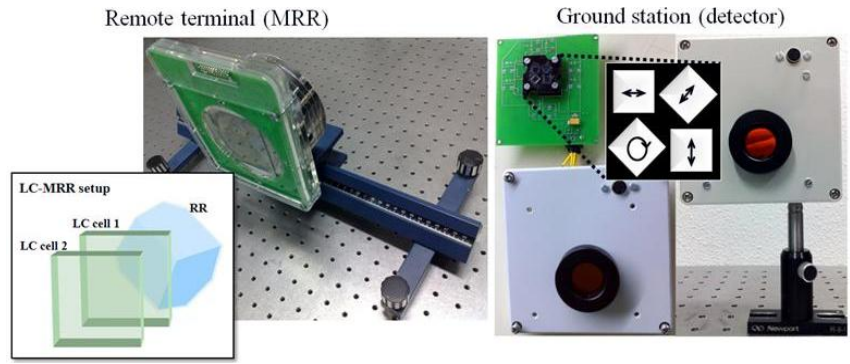

Figure 2. Liquid crystal-based MRR (left) and ground receptor (right).

Although the device (Fig. 2) had great advantages such as its low consumption and light weight, the slow response time of the liquid crystals has proven to be a handicap difficult to overcome, reaching speeds only in the order of tens of kHz and data rates under 100 kbps.

## 2.2 MEMS-based modulating retroreflector

The active device in the MRR finally designed is an MEMS modulator developed by Boston Micromachines Corporation (BMC), and it will be implemented in future works. This modulator is a reflective diffraction grating with controllable groove depth. The retroreflector is made up by three mirrors, being the MEMS modulator one of them (Fig. 3), OOK modulating a continuous laser beam by switching between an energized-diffractive state and an unpowered flat-mirror state.

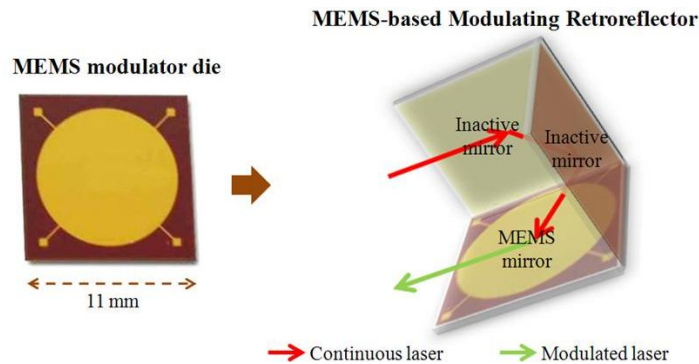

Figure 3. MEMS-based Modulating Retroreflector device.

The modulator is fabricated on a conductive substrate that works as one electrode of an array of elongated electrostatic actuators. The mirror surface acts as the other electrode, which is manufactured using MEMS technology and is suspended and electrically isolated from the substrate by an array of anchor supports. With the application of a voltage between the two electrodes, the actuators experience deflection, thus corrugating the mirror surface (Fig. 4). The electronics that drives the MRR will involve a challenge, since the modulator needs over 100V to achieve a 100% contrast (although such a high contrast is not needed, as the next paragraph explains) and this high-voltage switching has to be very fast in order to reach the goal speed of >1Mbps, while keeping a low power consumption (expected to remain below 100mW).

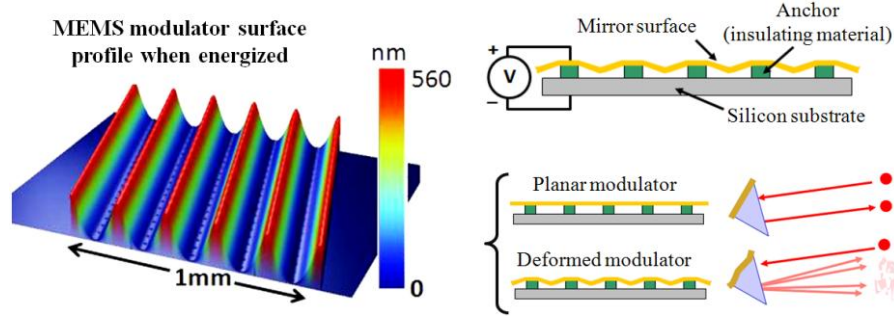

Figure 4. MEMS optical modulator surface profile and behaviour when energized (from [10]).

A trade-off exists between modulation contrast and dynamic response, so several devices will be tested, from 160µm to 200µm of actuator pitch, throughout the MRR development, in order to optimize the communications performance. Other improvements will also be addressed, such as different levels of vacuum environments during MRR fabrication so that the squeeze film damping effect can be minimized and the speed maximized.

## 3. GROUND STATION DESIGN

In the Fig. 5, a block diagram of the ground station is shown. In the following sections, a more detailed description of the different subsystems will be made.

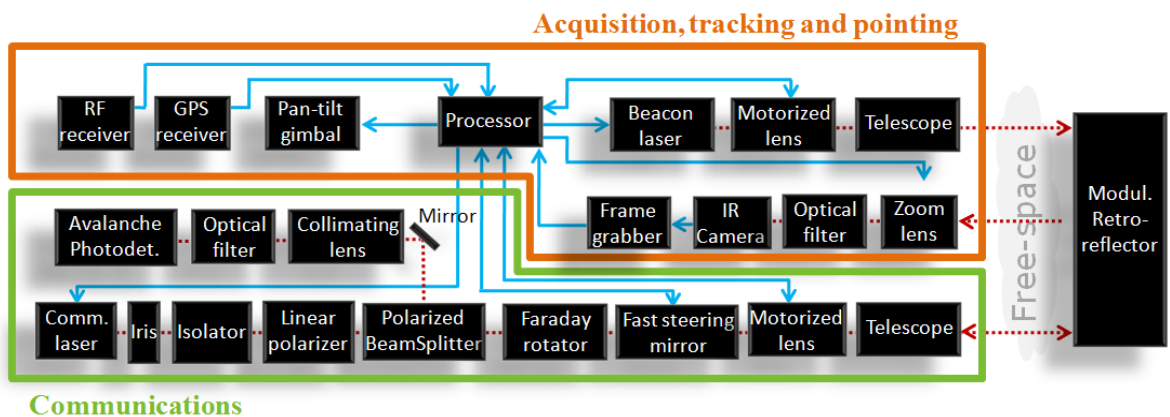

Figure 5. Optical ground station block diagram.

### 3.1 Acquisition, tracking and pointing

The ATP subsystem is a key part of the ground station, since the remote terminal is airborne, with a wide range of accelerations and speeds. It is based on a two-axis gimbal for tracking and coarse pointing, and a fast steering mirror (FSM) for fine pointing.

The most widely used strategy in lasercom is a beacon-based ATP technique [12]. However, a beacon system is not suitable onboard the UAV, due to weight constraints. Nevertheless, a beacon-based system, but with the laser in the ground station, has been designed taking advantage of the retroreflector system already installed onboard the UAV, with the advantages of a beacon-based system and with no extra payload. This strategy allows an optimal use of the MRR, acting as a communication terminal as well as a beacon reference for the ATP subsystem.

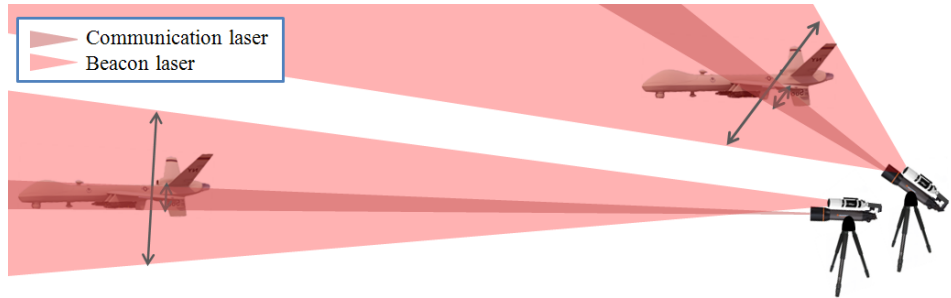

Figure 6. Beam divergence control to ensure a constant beam width regardless of the distance.

The designed strategy for the ATP system is as follows: the pan-tilt gimbal continuously tracks the trajectory of the UAV, aided by its GPS position captured by RF, and transmits an eye-safe 1550 nm beacon laser with a beam width wide enough to make up for the GPS error, guaranteeing that the UAV is always illuminated by the infrared beacon. This spot is estimated to be about 5 meters regardless of the distance, because of the divergence control described below. The beacon laser has a wavelength of 1550 nm and an optical power of 20 W, so that the intensity reaching the MRR is always 0.4 mW/cm$^2$. The gimbal is manufactured by FLIR Motion Control Systems and is a high-speed (up to 100°/s) system, capable of positioning a heavy load (<30 kg) with a resolution of 0.0064°.

Based on the retroreflector principle, the reflection of the beacon can be constantly monitorized through an IR camera in order to determine the exact position of the UAV. Fig. 7 shows a result of such a reflection detected by the IR camera. This was a laboratory experiment and it was made under distances and retroreflector sizes scaled to a realistic outdoor scenario (see section 4). As the distance between the two terminals is known through their GPS coordinates and its variability is high, a variable zoom lens will be needed in order to optimize the signal-to-noise ratio received by the IR camera, as well as a narrow optical filter to reject the wavelengths that differ from the one of the beacon. Considering the camera specifications, an initial maximum distance of 1,000 meters and a field of view of about 5 meters in the UAV, a variable focal distance between 10 mm and 100 mm is designed for the zoom lens.

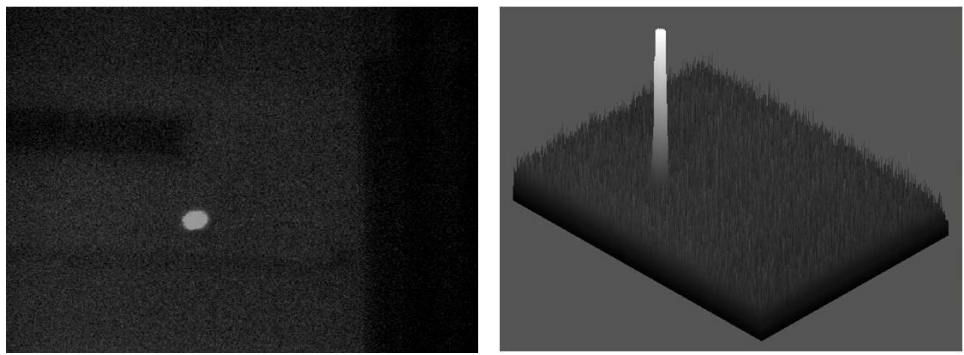

Figure 7. Image of the laser beacon reflection in the MRR (left) after processing with the IR camera in a laboratory experiment scaled to an actual scenario. The intensity level comparison of the detected pixels with the rest of the scenario is also shown (right).

The communication laser, with a much smaller (>20 times) beam width than the beacon laser (Fig. 6), is transmitted through a telescope controlling its fine position with the FSM, that can achieve a maximum angular resolution of 2 µrad. The communication laser beam reaching the MRR is about 10 cm, with an optical power of 100 mW. Fine pointing is based on a one-to-one correspondence between the position of the beacon image in the camera focal plane and the movement of the FSM that is needed to illuminate the MRR with the communication laser. This strategy makes it possible to maintain a real-time fine pointing with the UAV in a continuous mode.

## 3.2 Laser beam- divergence control

Since the UAV-to-ground station distance may vary greatly, a beam-divergence control has been designed to produce the same beam width reaching the UAV regardless of the instantaneous distance. This control is based on a converging lens travelling along a linear stage by means of a stepper motor, and a fixed 10 cm lens at the end of a telescope mount. The movements of the lens manage to defocus a laser that is collimated in the initial position (in which the two lenses make up a telescope), producing any desirable width beyond the diffraction-limited one (Fig. 8).

The laser-beam width reaching the UAV needs to be designed carefully, as there is a trade-off between optical power and spatial coverage. The beacon laser has to be wide enough to make up for the GPS-position uncertainty in order to illuminate the retroreflector continuously. Regarding the communication laser, this width has to be small enough to achieve a minimum signal-to-noise ratio to close the link, taking into account the way back from the retroreflector, but big enough to illuminate the MRR continuously making up for the movements of the UAV that the FSM cannot correct in time.

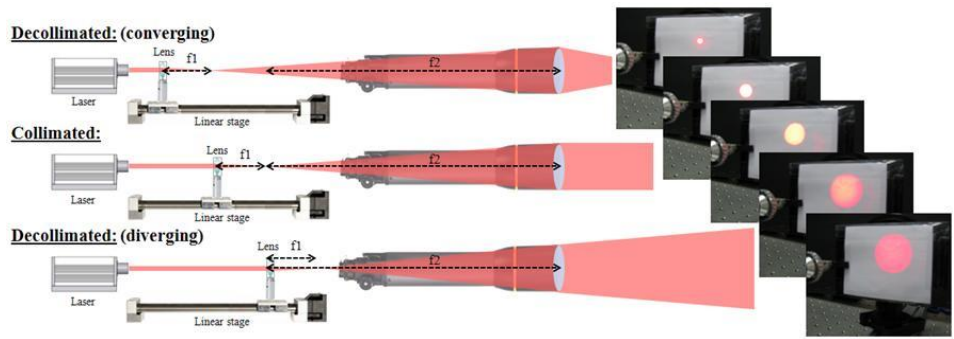

Figure 8. Laser beam-width control.

In the design of this beam width, it is also important to consider the FSM range of allowed angular laser movements before leaving the telescope and the minimum divergence achievable in the case of the largest link distance. Considering an initial maximum distance of 1,000 meters, according to the diffraction limit of an aperture at 1550 nm, a minimum beam width at the telescope would be 2 cm if the goal is to achieve coverage of ~10 cm reaching the remote terminal. It is important to note that in the distance range of an UAV, the beam width reaching the airborne is the result not only of the laser divergence, but also of the initial aperture (which is the beam width at 0 meters), negligible if the distance is longer (Fig. 9), as it is the case in most of the free-space lasercom links.

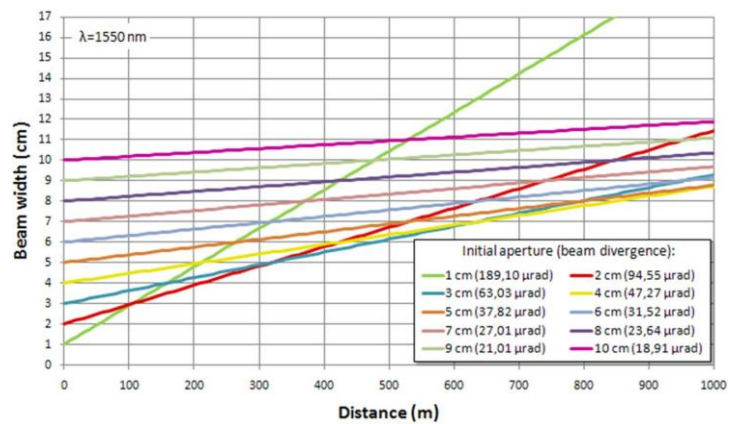

Figure 9. Beam width vs distance for a 1550 nm laser.

An important advantage of the designed beam-width control is that it is capable not only of increasing the divergence, but also of decreasing it, so the diffraction limit becomes a less restricting limit as it is possible to compensate a small beam-width leaving the telescope with a slightly converging decollimation in order to keep a very low divergence. This allows reducing the minimum beam-width under the 2 cm calculated before, which offers a bigger FSM movement range. This is possible because the beam-width reaching the UAV is constant for a given distance, so the FSM range is determined by the beam-width leaving the telescope.

In the designed system, the starting laser beam-width is 0,7 cm (after a beam expansion of 3,5× from the output of the laser). Considering an initial range of ground station-UAV distance between 10 and 1,000 meters, a 1 cm beam-width leaving the telescope and a 10 cm beam-width reaching the UAV, a divergence between 0,01° and 0,52° is needed. This can be achieved by a 100 mm-focal-distance converging lens travelling along a 12 cm linear stage with a 660 mm-focal-distance primary lens.

### 3.3 State of polarization discrimination

An optical communication link based on the MRR scheme results in an optical axis that is the same for the transmitted and the received beam, assuming that the beam width is bigger than the retroreflector aperture. Otherwise the returning beam would have an unknown offset from the interrogator optical axis, as it would depend on which of the three mirror faces of the retroreflector the beam passes first.

The simplest solution to enable the reception of the laser on its way back is to shift the ground receiver out of the optical axis, but an important part of the returning light is lost this way, as the center of the gaussian power distribution would fall on the interrogating laser axis.

The solution proposed in this paper is creating two different optical paths within the same physical path. The simplest way to receive the laser on the way back in the same physical path is by means of a 50-50 beam splitter, by deflecting the laser on its way back towards a photodetector placed in a 90 degrees direction (Fig. 10). However, this method has an important drawback, as 50% of the optical power is lost in each pass. Since the wavelength is always the same, a spectral beamsplitter cannot be used.

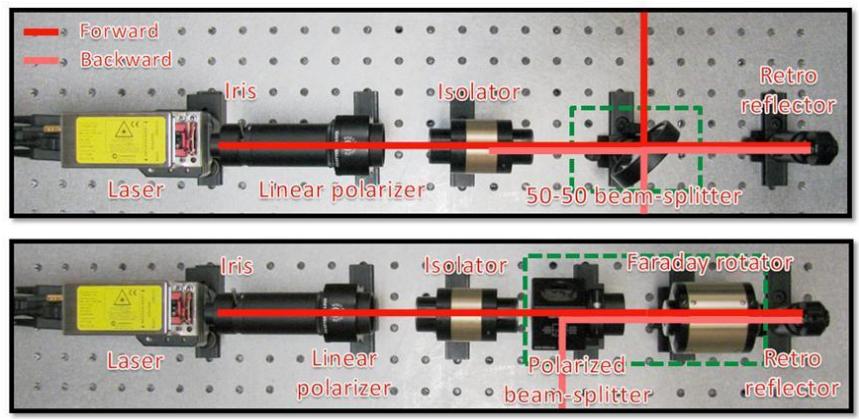

Figure 10. Trajectory discrimination setup (up) and polarization discrimination setup (down).

The proposed technique makes it possible to minimize the lost power by transmitting in a linear polarization and receiving in the orthogonal polarization on the way back. This can be accomplished with a setup consisting of a Faraday rotator –since 45° linear polarization is rotated in each of the two passes– and a polarized beam splitter –which deflects the beam to a 90 degrees direction when aligned with the orthogonal polarization.

In the setup of Fig. 10, a N-BK7 linear polarizer was added to impose a >20,000:1 extinction ratio, which is a key feature in order to optimize the performance of this setup, where a pure linear polarization achieves the best results. The isolator was used just for the sake of protection of the laser system from any residual light on the way back, essential in the trajectory discrimination setup, but not indispensable in the polarization discrimination setup, as long as the polarized beam splitter is well aligned with the laser beam. A gain of ~7dB has been measured this way compared with the 50-50 beam splitter setup. The most significant part of this gain comes from the fact that there is no loosing of a 50% of the power in each of the passes, which makes up 3dB+3dB of gain, and the rest is gained thanks to the use of better coated optics.

## 4. CONCLUDING REMARKS AND FUTURE WORK

The design of a lasercom system has been addressed and the ground station has been partially developed (Fig. 12). The full development of the ground station, including the integration of all the subsystems, involves the main workload in the project and is being done at the moment. A laboratory test bed is being prepared including a trial for every part of the system, as well as a battery of tests of the final integrated system. An elevated rail layout has been built in the laboratory, along which a robot will be travelling. In the first tests, this robot will be carrying the MRR, and its movements will be equivalent to the ones of a distant UAV. A miniature retroreflector will make it possible to simulate a long-distance experiment in the fine-pointing test. A series of field tests are also programmed, involving the use of a small radio-controlled airplane to begin with, and eventually an actual UAV for the final tests at the end of the project, the use of which is being negotiated.

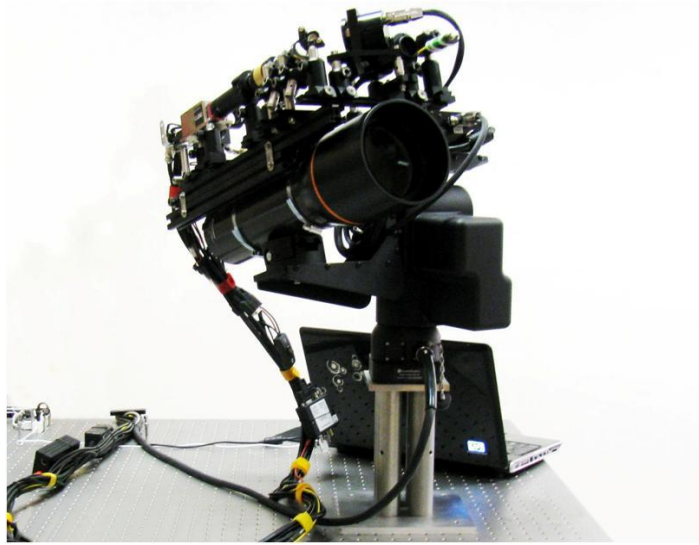

Figure 12. Current state of the partially-developed ground station.

Future system updates will include higher data rates using fiber optics technology at ground station, atmospheric turbulence mitigation with aperture averaging and adaptive optics, active inertial stabilization of the ground station to make it suitable for mobile platforms and a plan to integrate laser power transmission into the system, taking advantage of the ATP subsystem already developed.

## ACKNOWLEDGMENT

This work is supported by INSA (*Ingeniería y Servicios Aeroespaciales S.A.*), within the SINTONIA project (*Sistemas No Tripulados Orientados al Nulo Impacto Ambiental*), CENIT-E 2009 program, funded by CDTI (*Centro para el Desarrollo Tecnológico Industrial*, grant. CEN-20091001), and Comunidad Autónoma de Madrid: CAM (grant. nº S2009/ESP-1781, FACTOTEM-2).